# Simulation of a Compton-based detector for low-dose high-resolution time-of-flight positron emission tomography

Kepler Domurat-Sousa[1,2], Cameron M. Poe[1], Maya S. McDaniel[2], Eric Spieglan[1], Joao F. Shida[1], Evan Angelico[1], Bernhard W. Adams[3], Patrick J. La Riviere[4], Henry J. Frisch[1], and Allison H. Squires[2,5*]

1. Enrico Fermi Institute, University of Chicago, Chicago, IL USA
2. Pritzker School of Molecular Engineering, University of Chicago, Chicago, IL, USA
3. Quantum Optics Applied Research, Naperville, IL, USA
4. Department of Radiology at the Pritzker School of Medicine, University of Chicago, Chicago, IL, USA
5. Institute for Biophysical Dynamics at the University of Chicago, IL, USA
   * To whom correspondence should be addressed: asquires@uchicago.edu

## I. ABSTRACT

**Background**: Two major challenges in time-of-flight positron emission tomography (TOF-PET) are low spatial resolution and high radioactive dose to the patient, both of which result from limitations in detection technology rather than fundamental physics. To address these challenges, a new type of TOF-PET detector employing low-atomic number (low-Z) scintillation media and large-area, high-resolution photodetectors to record Compton scattering locations in the detector has been proposed, but the minimum technical requirements for such a system have not yet been established.

**Purpose:** The purpose of this study is to computationally assess a proposed low-Z detection medium, linear alkylbenzene (LAB) doped with a switchable molecular recorder, and to calculate reasonable specifications of energy, spatial, and timing resolution for a next-generation TOF-PET detector.

**Methods:** We validated and used a custom Monte Carlo simulation of full-body TOF-PET with a low-Z detector within the TOPAS Geant4 software package, producing simulated data for varying energy, spatial, and timing resolutions of the detector. We developed an algorithm to produce the most likely lines of response and reconstructed images by simple addition of source distribution probabilities. We evaluated the effect of changing detector specifications on TOF-PET sensitivity and on the resolution and contrast-to-noise ratios of images of standard phantoms. Using reasonable specifications, we simulated low-Z TOF-PET for a brain phantom with a lesion at significantly reduced doses.

**Results:** Our likelihood-based identification of pairs of first interaction locations in the simulated detector identifies 87.1% of pairs with zero or negligible error, and correctly rejects 90% of all in-patient scatters. As expected, changing the detector energy, spatial, and timing resolutions tuned the simulated sensitivity, resolution, and contrast-to-noise. We found that even modest detector specifications (energy resolution 1 keV/switch, spatial resolution 1 mm, time resolution 500 ps FWHM) produce TOF-PET sensitivity of ~66.7% and PSF width 4.6 mm with clear contrast. A detector with these specifications provides a clear image of a brain phantom simulated at less than 1% of a standard radiotracer dose.

**Conclusions:** A low-Z TOF-PET detector with reasonable technical specifications could substantially lower dose requirements while maintaining or improving image resolution and contrast. While the engineering details of such a detector are yet to be experimentally explored, the present work supports the feasibility of this Compton-based TOF-PET detection method which could ultimately enable expanded access and new clinical applications for TOF-PET.

**Short title:** "Compton-based TOF-PET"

## II. INTRODUCTION

Time-of-Flight Positron Emission Tomography (TOF-PET) provides medical insights into the metabolic and biochemical function of tissues and organs by mapping relative uptake of a positron-emitting radiotracer. TOF-PET imaging traditionally relies on detecting coincident pairs of 511 keV photon arrivals recorded by scintillation crystals arrayed around the subject. Despite its biological specificity, the major drawbacks of TOF-PET remain (1) low detector sensitivity such that only 1-2% of emitted photons are detected in standard (non-full-body) commercial PET systems,[1] setting the minimum radioactive exposure for patients, the minimum imaging time, and the geographic accessibility of PET; and (2) image resolution that is limited by the use of a crystal array detector, which discretizes the observed locations of photon interactions and limits the minimum measurable size of features such as cancer metastases.[2,3] These limitations arise from technical constraints of the detector technology, such as photoelectric cross section, crystal size and scintillation efficiency, rather than underlying physical principles governing positron annihilation, photon emission, and photon-detector interactions.[4]

State-of-the-art TOF-PET imagers implement a near-optimized version of this traditional detection approach.[1,5–9] Development of full-body PET systems,[2,10,11] combined with optimization of detector timing precision,[12–16] trajectory resolution, [17,18] detector material, [19–21] and image reconstruction algorithms[22,23] have enhanced the efficiency with which pairs of photons are detected to sensitivities above 15% and improve image resolution to < 3 mm.[24] Yet future TOF-PET development may require order-of-magnitude advances in timing precision or a new detector paradigm to overcome the sensing limitations and cost tradeoffs of scintillation crystal arrays. In this context, we have been exploring TOF-PET scanner designs based on inexpensive low atomic number (low-Z) scintillating media.[25–30]

The proposed low-Z TOF-PET detection approach, shown schematically in Fig. 1, exploits Compton scattering rather than the photoelectric effect as the fundamental photon-detector interaction. A low-atomic number (low-Z) liquid scintillation medium is used instead of high-atomic number scintillation crystals, making Compton scattering the predominant interaction [31]. This approach is similar in concept to a "Compton camera",[32,33] but allows for detection of an arbitrary number of scatters and geometries because it uses an entirely different implementation. Upon scattering in the low-Z detector, a recoil electron and scintillation light are generated. With each successive Compton scatter, the photon continues to lose energy until it either leaves the detector or is absorbed. The observed geometry and energies of a chain of Compton scatters, together with its counterpart across the detector, may be used to statistically infer which event occurred first in each chain (inset, Fig. 1). Lines-of-response (LORs) are determined using these likely pairs of first scatters across the subject, from which the original distribution of radiotracer in the subject may be estimated. The two major proposed advantages of such a low-Z TOF-PET detector are (1) that sensitivity could be arbitrarily raised by increasing the depth of a low-cost liquid detection medium, and (2) localization of interactions would be continuous rather than discretized, removing a crucial barrier to achieving ultra-high-resolution TOF-PET images. However, a quantitative evaluation of the technical feasibility of this approach has not been done.

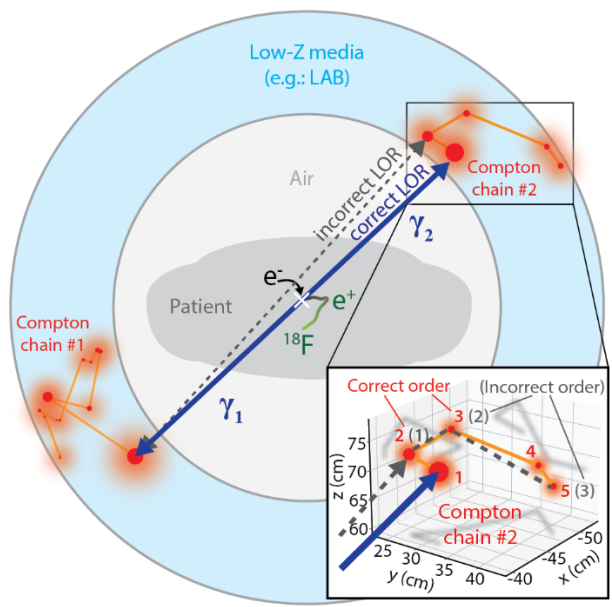

**Fig. 1.** Schematic of proposed TOF-PET detection scheme using the first Compton scatter location to determine a line of response (LOR). Low-Z detection media such as linear alkylbenzene (LAB) containing a switchable reporter such as a photoswitchable fluorophore surrounds the patient. A positron (green e⁺) originating from the decay of a radiotracer (here: $^{18}F$) diffuses (green track) and annihilates with an electron (black e⁻, white x). Each of the two resulting back-to-back photons (dark blue) interacts with the LAB via a chain of Compton scattering events (red/orange; circle area is proportional to recoil electron energy). The correct line-of-response (LOR) connects the first Compton scatter of each chain; mis-identification of the first interaction will result in an incorrect LOR (gray dashed line). *Inset:* Different possible orderings of Compton scatters in one simulated chain (shadows show 2-D projections) are proposed and selected using a likelihood-based figure of merit (FOM); here one incorrect ordering is shown (gray dashed line; FOM = 20.88) along with the true ordering (orange / blue; FOM = 0.922).

Here we present Monte Carlo simulations of full-body low-Z TOF-PET using the TOol for PArticle Simulation (TOPAS) Geant4 software package,[34,35] with which we evaluate the feasibility, technical requirements, and potential advantages of such a low-Z TOF-PET detector approach. We simulated data of chains of Compton scatters from photons interacting with the detector for varying energy, spatial, and timing resolutions. We developed and evaluated an algorithm to analyze these chains of Compton scatters to determine the most likely lines of response and to screen out in-patient scatters and unpaired interactions with the detector. Using images reconstructed by simple addition of the line of response source distribution probabilities, we evaluated the effect of changing detector specifications on TOF-PET sensitivity, on the resolution, and contrast-to-noise ratios of images of standard phantoms. Using specifications with reasonable performance, we simulated low-Z TOF-PET for a brain phantom with a lesion at doses reduced by orders of magnitude to illustrate one potential high-impact use case.

# III.  METHODS

## A.  TOPAS Simulation of TOF-PET

We constructed and parameterized a simulation of a whole-body TOF-PET scanner using TOPAS 3.8[34] running the Geant4 physics modules "g4em-standard opt4" and "g4em-penelope"[36–40]. The cylindrical detector (2 m long; bore diameter 90 cm; radial thickness 30 cm) is filled with high purity linear alkylbenzene (LAB) (density 0.860 g/cm³, composition 87.86% H, 12.14% C[41]; mean excitation energy 59.4 eV[42]), with an air core. Intrinsic radioactivity of detector media is expected to be negligible and was not simulated.

Different $^{18}$F imaging sources can be positioned within the bore of the detector; in this work we used the Derenzo geometric phantom[43], the XCAT human phantom[44], a positron point source, and the NEMA NU 2-2018 sensitivity source.[45] The per volume activity for each phantom is benchmarked to typical activity in a patient with a "standard dose", calculated as 5 MBq/kg[46] and a baseline imaging time of 10 minutes. All simulations in this work were performed with a reduced dose, set at or below 1/100 of the standard dose. Simulations include in-patient scattering.

TOPAS generates a set of $^{18}$F positrons throughout the source volumes, which are passed to Geant4 for propagation. Geant4 handles positron diffusion and annihilation, generating pairs of photons that propagate through the simulation volume. We wrote a custom C++ scoring extension for TOPAS to record events in the detector volume. The minimum step size was set to 10 μm. Simulations were run on a 14-thread processor (Intel Xeon CPU E5-2620v4 at 2.10 GHz), with runtimes between 15 and 70 hours for 83 million positrons, depending on the complexity of the phantom geometry. Ground truth output data include times and Cartesian positions of interactions, the energy balance of each interaction, and all particle identities. All custom code and supporting documentation are available for download online[47].

## B.  Simulation of Detector Resolution

Random noise was applied to the ground truth simulation data according to the detector's energy, timing, and spatial resolution parameters. Error in energy measurements arises from uncertainty in the magnitude of a recoil electron's effect on the detection medium. All added noise was Gaussian-sampled. Assuming a two-state molecular recorder, the efficiency $k_E$ with which the recoil e⁻ of energy $E$ produces switched molecules, $N(E) = E/k_E$ (units keV/switch), determines the back-propagated uncertainty in electron energy, $\sigma_E$, using the uncertainty of counting $N$ switched molecules, $\sqrt{N}$, according to

$$\sigma_E(E) = k_E \sqrt{N(E)} = \sqrt{k_E \cdot E}, \tag{1}$$

where $N(E)$ is rounded to the nearest integer. If the resulting noise-corrupted $E$ happens to be negative (possible for small $N(E)$ events), the event is discarded. Spatial uncertainty, $\sigma_x$, is directly applied to ground truth scattering locations in three dimensions. Timing uncertainty as a Full Width Half Maximum, FWHM$_t$, is directly applied to ground truth interaction times. For the purposes of the following analyses we ignore pile-up; we assume that the records corresponding to each chain of Compton scatters can be correctly grouped. This might be experimentally achieved, for example, using coincidence timing of initial light from each scatter.

The default resolutions for simulation are set as follows: The energy resolution is set by $k_E$ with a default value of 1 keV/switch. The default spatial resolution of scattering locations, $\sigma_x$, is a 1 mm standard deviation. The default timing resolution, FWHM$_t$, is conservatively set to 500 ps. Variation of these parameters is studied in *IV.C. Effect of Tuning Detector Parameters*.

## C. Statistical Determination of Lines-of-Response

Simulated detector data are used to reconstruct TOF-PET images based upon the aggregate information from many individual lines of response (LORs). For low-Z detection, the line of response for each annihilation event is drawn between the first Compton scattering location for the back-to-back photons. Image reconstruction therefore first requires correct ordering of each Compton chain, and rejection of in-patient scatters and unpaired events. While a detector may provide timing information, it is not strictly necessary; here we develop an ordering and screening algorithm based solely on the geometry and energy of scattering interactions.

We developed an algorithm to screen out in-patient scatters and determine the most probable ordering (and therefore first scatters) in each pair of Compton chains using the overall geometry and measured energies of each set of scatters. For each individual event $i$ causing an apparent change in trajectory, $\theta_i$, of a photon and producing a recoil electron with energy $E_i$, the expected incoming photon energy, $E_{\gamma i}^{\text{Comp}}$, is given by

$$E_{\gamma i}^{\text{Comp}} = \frac{E_i + \sqrt{E_i^2 + \frac{4E_i m_e}{1 - \cos \theta_i}}}{2}, \tag{2}$$

where $m_e$ is the rest mass of the electron.

An alternative estimate of the incoming photon energy, $E_{\gamma i}^{\text{chain}}$, can be calculated based on the preceding series of recoil electron energies $E_j$, along with the expected photon energy of 511 keV:

$$E_{\gamma i}^{\text{chain}} = 511 \text{ keV} - \sum_{j=1}^{i-1} E_j, \tag{3}$$

where $E_j$ is the observed energy at scatter $j$.

Propagation of error gives the uncertainty in each estimate of $E_{\gamma i}$ as calculated using $k_E(E)$ and the uncertainty in angular redirection, $\sigma_\theta(\vec{x}_{i-1}, \vec{x}_i, \vec{x}_{i+1}, \sigma_x)$, which is calculated from scatter locations and the spatial uncertainty $\sigma_x$. Given these uncertainties, we generate a figure of merit for each individual scatter in a proposed chain order, FOM$_i$, using a two-tailed Z-score:

$$\text{FOM}_i = \frac{\left| E_{\gamma i}^{\text{chain}} - E_{\gamma i}^{\text{Comp}} \right|}{\sqrt{\left( \delta E_{\gamma i}^{\text{chain}} \right)^2 + \left( \delta E_{\gamma i}^{\text{Comp}} \right)^2}}, \tag{4}$$

where $\delta E_{\gamma i}^{\text{chain}}$ is the uncertainty in $E_{\gamma i}^{\text{chain}}$, and $\delta E_{\gamma i}^{\text{Comp}}$ is the uncertainty in $E_{\gamma i}^{\text{Comp}}$. Large FOM indicates a large difference between $E_{\text{chain}}$ and $E_{\text{Comp}}$, which indicates incorrect ordering (giving incorrect $E_{\text{Comp}}$ values) or an incorrect starting photon energy (giving incorrect values of $E_{\text{chain}}$) signifying a photon that in-patient scattered. Therefore, we reject all FOM greater than a cutoff value of 1.3*($N_{\text{comp}}$) where $N_{\text{comp}}$ is the number of scatters. This cutoff was selected to optimize the balance of detector sensitivity and rejection of in-patient scatters.

Possible chain orders are tested in a recursive tree search, keeping only the current minimum FOM solution and terminating search branches above the FOM cutoff. Unpaired chains where only one photon created a detector interaction are discarded. Pairs of chains where one or both photons undergo significant in-patient scattering fail the FOM test.

## D. Image Reconstruction

The LOR defined by locations and timing of each pair of first scatters is a needle-shaped 3-D Gaussian probability density describing the location of each annihilation. A TOF-PET image can be reconstructed based on the collective spatial distribution of all LORs. Filtered back-projection (FBP) and other common TOF-PET reconstruction algorithms are designed for discretized LORs and traditional detectors. Here, LOR positions and orientations for a low-Z detector are continuous; a sophisticated reconstruction algorithm using appropriate weights would require a fully characterized active medium, a realistic optical design, and would take full advantage of the correlated spatial and temporal data. Therefore, in this work we reconstruct images by summation of probabilities for all LORs over a voxelized volume slice (22 cm x 22 cm x 0.5 mm with 0.125 mm$^3$ cubic voxels). As an approximate correction for image-to-image variation in LOR width, each image is normalized by the integrated transverse profile at the center of an LOR perpendicular to the field of view. All reconstructed image slices shown in this work are taken perpendicular to the detector axis, centered on the origin.

## E. Sensitivity, Spatial Resolution, and Contrast-to-Noise (CNR) Criteria

Two fundamental criteria for evaluating TOF-PET performance are the efficiency in detecting emitted photon pairs, or *sensitivity*, and the width of its point spread function, or *spatial resolution*. Here, for comparisons of various detector resolution parameters we calculate sensitivity, $S$, directly from simulated data of a point source in vacuum placed at the origin, using the ground truth number of pairs of photons that interact with the detector, $n_{pair}$, compared to the total number of positrons simulated, $N_{e+} = 10^6$ positrons, according to

$$S = \frac{n_{pair}}{N_{e+}} \tag{5}$$

This criterion is based on the NEMA NU 2-2018 definition of TOF-PET sensitivity,[45] substituting a point source for the line source so that the same simulated data can be used to calculate the point spread function (PSF). The full width at half maximum (FWHM) and full width at tenth maximum (FWTM) of a transverse image slice through the reconstruction of this PSF define the spatial resolution per the NEMA NU 2-2018 criteria.[45] For the sensitivity as reported in section *IV.C. Performance of simulated low-Z TOF-PET* the NEMA NU 2-2018 procedure was followed in the simulation.

To evaluate the fidelity and quality with which reconstructed images represent the true radiotracer distribution within the subject, we also calculate the contrast-to-noise ratio (CNR) for well-established features of certain phantoms according to

$$CNR = \frac{\bar{p}_1 - \bar{p}_0}{\sqrt{\sigma_{p1}^2 + \sigma_{p0}^2}} \tag{6}$$

where $\bar{p}_1$ is the average value of pixels in the signal region, $\bar{p}_0$ is the average value of pixels in the background region, $\sigma_{p1}$ is the standard deviation in the signal region, and $\sigma_{p0}$ is the standard deviation of pixels in the background region. Rod CNRs are averaged by size group. See the *Supporting Information* for mask details and rod-by-rod CNR analysis.

## IV.   RESULTS

### A.  Simulation design and parameter selection

To record a chain of Compton scatters in the detector, energy from each recoil electron must locally alter the recording medium to indicate the position of the scatter. The extent of this change should correlate to and therefore report the energy of the recoil electron. The change must persist for sufficient time to record the interaction with high spatial and energy precision, and the medium should be resettable after each measurement. The recording mechanism could be thermodynamic, as in a bubble chamber, but chemical changes in molecular state or configuration could offer greater sensitivity and opportunities for external control: One such implementation would utilize photoswitchable fluorophores[48–50] that can be locally activated by energy from a recoil electron to switch from a native "dark" state to a fluorescence-capable state that may undergo many cycles of excitation and emission, enabling extended-duration recording of the location and number of switched molecules before returning to the dark state, as described previously[27] and depicted in Fig. 1. The simulations performed in the present work are agnostic to the specific recording mechanism, and instead take the efficiency of converting recoil electron energy into a molecular record as a simulated detector parameter for energy resolution.

Regardless of mechanism, a major advantage of a low-Z detection approach to TOF-PET is that estimation of the locations and energies of photon interactions in the detector can be decoupled from timing measurements, so that these aspects of detection may be independently optimized. The low cost and wide availability of low-Z detection media such as organic solvents means that the sensitivity of the detector may also be independently tuned to achieve high stopping power through altering the detector volume. In the following simulations, we employ LAB as the low-Z detection medium. LAB has excellent optical clarity at visible wavelengths, high flash point, is low-cost, and has relatively high viscosity at room temperature, limiting solute diffusion.

### B.  Simulation Validation

We next validated our simulation by comparing its output with well-established physical phenomena. Fig. 2a and 2b show the simulated distributions of drift distance and kinetic energy, respectively, of positrons at annihilation. The distributions agree well with published results: the FWHM of the positron drift distance is 0.35 mm, and the majority (98.4%) of positrons have identically zero kinetic energy upon annihilation.[51–53] Fig. 2c shows the distribution of initial distances traveled into LAB by photons prior to interaction, as well as the distance between the first and second scatters in LAB. The simulated initial distance corresponds well to the calculated $1/e$ stopping distance of LAB (12.05 cm).[31]

The scattering angles and the outgoing energies in each scatter are constrained by the electron-photon two-body kinematics. Fig. 2d shows the event-by-event relationship between the simulated scattering angle and the recoil electron energy for primary 511 keV photons. The Compton formula (dotted blue) fits the simulated data well. The inset shows a profile at a scattering angle of $\pi/4$. The profile width is due to modeling of electron binding and kinetic energy by the Penelope package,[54] and has a two-tailed exponential spread with a decay constant of 2.04 keV[55], matching the expected value for carbon.[55]

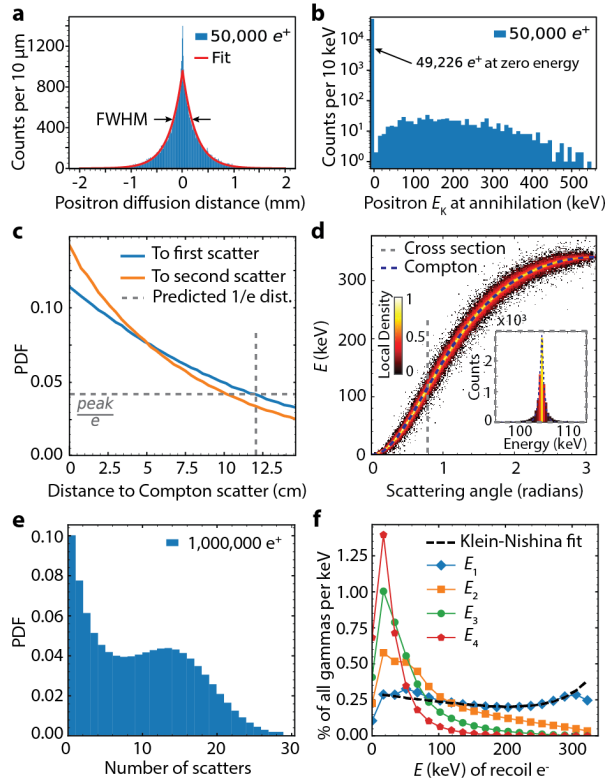

**Fig. 2.** Simulation validation and characterization. a) Histogram of positron diffusion distance prior to annihilation (blue) and fit (red), with FWHM 0.35 mm. b) Histogram of positron kinetic energy at annihilation (blue) is mostly identically 0 keV (98.4%). C) Photon distance into the LAB medium before first Compton scattering event (blue) and distance between first and second Compton scattering events (orange). The expected $1/e$ range for LAB (12.05 cm) is shown in gray dashed crosshairs. d) 2-D histogram of energy deposited in the first Compton scattering event ($E$, red-yellow heatmap) for each scattering angle, closely matches the theoretical prediction from the Compton relation (dotted blue). Inset: Cross-section at $\theta = \pi/4$, with double-sided exponential fit (dotted blue). E) Histogram of the number of Compton scattering events observed in each chain. f) Histograms of energy deposited in the first (blue / diamond), second (orange / square), third (green / circle) and fourth (red / pentagon) Compton scatters in a chain, respectively. A fit to the Klein-Nishina cross section is shown as a black dashed line.

The ground truth simulation results also benchmark the expected behavior of a low-Z TOF-PET detector: Fig. 2e shows the distribution of the number of scatters observed in Compton chains within the detector, averaging 10.1 scatters per chain, heavily skewed towards a lower number of scatters. Fig. 2f shows the fraction of all Compton scattering events at each recoil electron energy $E$. As expected, the recoil electron energy trends downward with successive scatters, and the distribution of first scatters is well-fit by the Klein-Nishina model (dotted black).[56]

### C. Performance of simulated low-Z TOF-PET

To determine if a low-Z TOF-PET detector could be used to faithfully reconstruct an underlying source distribution, we simulated a Derenzo phantom[57] and the NEMA NU 2-2018 line source sensitivity phantom[45] using the default simulation parameters ($k_E$ = 1 keV/switch, $\sigma_x$ = 1 mm, FWHM$_t$ = 500 ps) as described in *III. Methods*. A 1 mm thick slice from a simple reconstruction of a simulated Derenzo phantom at 1/100 dose is shown in Fig. 3a. All small cylindrical features are clearly visible with uniform intensity, and the background cylinder body is clear and even except for slight blurring at the outer edges due to the simple reconstruction approach. For the NEMA NU 2-2018 line source sensitivity phantom, the low-Z detector simulation results in 66.7% sensitivity (see *Supporting Information* Fig. 2 for attenuation curve). Even with the simple image reconstruction approach employed here, the high predicted sensitivity of the low-Z TOF-PET detector allows for clear imaging at a low dose.

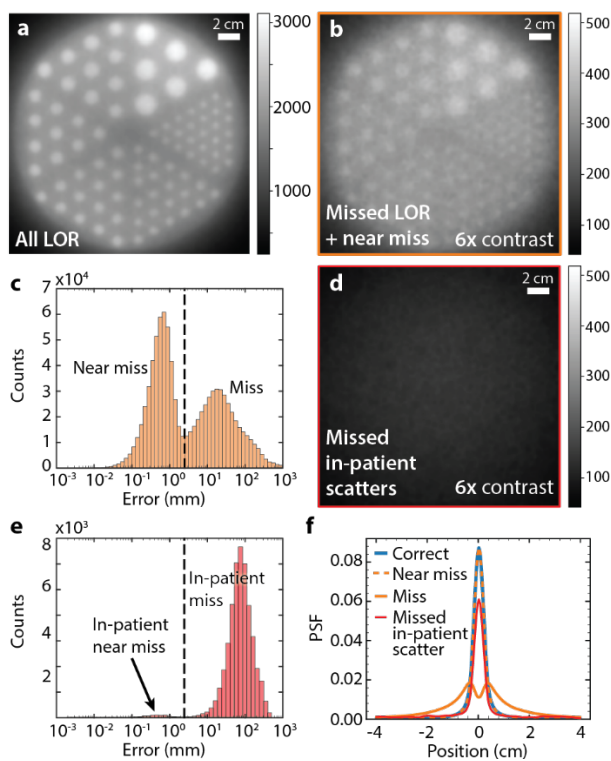

**Fig. 3.** Effect of incorrect Compton chain ordering on reconstructed image and LORs. a) Image reconstructed from all LORs ($3.69 \times 10^7$ LORs). b) Image reconstructed from all Compton chains that did not result in correct ordering, shown at 6x enhanced contrast as compared to 3a) ($1.10 \times 10^7$ LORs). c) Histogram of error between all incorrect LORs and ground truth for a point source with no in-patient scatters, divided into two populations "near miss" and "incorrect" according to the magnitude of error (dashed line set at 2 mm). d) A reconstruction of only the in-patient scattered LORs of the Derenzo phantom shown at 6x enhanced contrast compared to 3a. e) Histogram of error between all in-patient scatters that formed LORs and ground truth for a point source centered in a Derenzo phantom sized cylinder of water, divided into two populations "near miss" and "incorrect" using the magnitude of the error (dashed line at 2 mm). f) PSFs of correct LOR as compared to near misses, misses, and in-patient scatters LOR for a point source. FWHM of correct and near miss LOR is 4.6 mm.

Because a low-Z detection approach requires likelihood-based ordering of a Compton scattering chain, in which events can be separated by many centimeters, mis-identified first scatters can cause LOR and reconstruction errors that would not be present in standard TOF-PET. As with standard TOF-PET, any in-patient scatters that are not rejected will also form LORs that reduce image quality. Therefore, the image in Fig. 3a contains three categories of LOR: Correctly identified first scatter pairs ("Correct": $2.587 \times 10^7$ LORs; 70.1%), LORs where one or both first scatters were incorrectly identified even though neither photon underwent in-patient scattering ("Missed LOR": $4.094 \times 10^6$ LORs; 11.1%), and LORs reconstructed from instances where one or both photons underwent in-patient scattering but the FOM was below the cutoff ("Missed in-patient scatters": $6.940 \times 10^6$ LORs; 18.8%).

To determine the effect of incorrect LORs on the reconstructed image, we constructed an image from only the missed LORs, shown in Fig. 3b. Surprisingly, this image is still recognizable as a Derenzo phantom, indicating that some incorrect LORs still contain valuable information. Indeed, a distribution of the miss distance for all incorrect LORs, shown in Fig. 3c, is bimodal. LORs that miss the true annihilation location by several centimeters are a result of general failures of the chain ordering algorithm as expected. The second population, which we term "near miss", falls within 2 mm of the correct location of annihilation. Approximately half of missed LORs are near misses, accounting for the substantial information content of this LOR category. We found that the near-miss LORs tend to arise from Compton chains with one of two specific features: The first type of feature is a very low energy first scatter (10 keV or less) followed by a larger second scatter. Due to the very small first scatter, the second scatter tends to be close to the original trajectory, and therefore produces a near-miss LOR if it is mis-identified as first. Alternatively, if any scatter that is mis-identified as first is sufficiently close to the true first scatter, the LOR will be a near miss. Such close proximity of another scatter to the first scatter limits the discrimination of the ordering algorithm, but without a meaningful loss in fidelity.

For the Derenzo phantom, the Compton chain FOM algorithm for identifying first scatter pairs rejects 90% of all in-patient scatters. The in-patient scatters have photon energies of less than 511 keV when they enter the detector, so the FOM fails to find likely physical solutions for their recorded trajectories. See Supporting Fig. 3 for the distribution of all photon energies that resulted in an LOR. Nevertheless, because the rate of in-patient scattering is high, the missed 10% of inpatient scatters might significantly impact image quality (for the Derenzo phantom, 67.7% of all photon pairs in the detector underwent in-patient scattering). To check this, we reconstructed the Derenzo phantom image using only the LORs corresponding to missed in-patient scatters, as shown in Fig. 3d. It is clear from this image that LORs from missed in-patient scatters produce a low frequency background due to large and random deviations from the annihilation locations. Although the histogram of LOR error distances for missed in-patient scatters shown in Fig. 3e is bimodal, very few of these LORs are near misses.

The PSFs resulting from each type of LOR are shown in Fig. 3f. The correct and near-miss LORs have a FWHM of 4.6 mm, and the broad tails of the miss and in-patient scattered LORs can be seen. The sharp peak in the in-patient scatter PSF is from the near-miss population. This peak appears despite being a small fraction of in-patient scattered LORs (0.2%) due to the near-miss LORs being tightly spaced while the other in-patient scattered LORs are spread over a large volume. For the correct, near-miss, and miss LORs a point source of positrons with no attenuation was used. For the in-patient scatters, a point source located in the center of a cylinder of water the size of the Derenzo phantom was used to determine the PSF.

## D. Effect of Tuning Detector Parameters

To place upper and lower bounds on the technical requirements and to better understand the potential advantages and limitations of low-Z detectors for advanced TOF-PET, we simulated a range of detector resolution parameters, applied as described in *III. Methods*. For each parameter, we determined the effect on (1) image spatial resolution as measured by the FWHM and FWTM of the PSF for a point source, (2) detector sensitivity, including a breakdown of correct, incorrect, and near-miss LOR percentages, and (3) the average CNR for rods in the Derenzo phantom at 1/100 dose. For each set of parameters, we also display the best Derenzo image.

Energy resolution is controlled by $k_E$, the efficiency of switching (keV/switch). We tested $k_E$ values of 1, 5, 10, 50, and 100 keV/switch (default: 1 keV/switch). Fig. 4a-c show the effect of energy resolution on the image spatial resolution, the detector sensitivity, and the CNR for Derenzo rods. Changes in $k_E$ do not lead to a substantial change in image spatial resolution (Fig. 5a) but decreasing energy resolution (higher $k_E$) leads to substantial losses in detector sensitivity, with many scatters no longer being correctly identified (Fig. 5b). The decreasing sensitivity also causes loss of CNR in the Derenzo phantom shown in Fig. 4c. The best Derenzo image, shown in Fig. 4d, results from the best energy resolution (1 keV/switch).

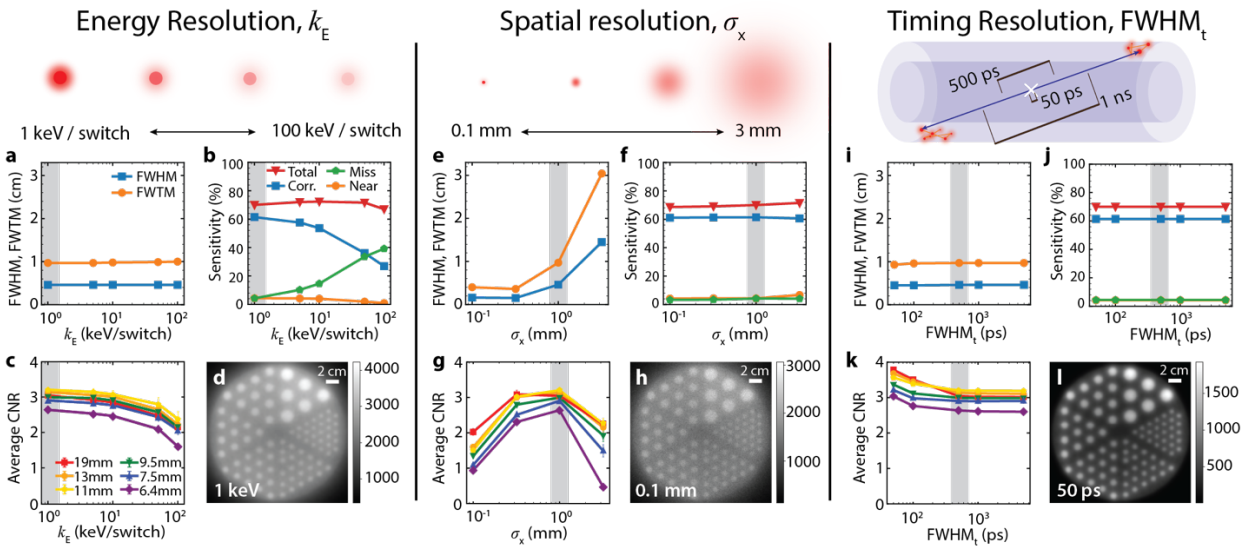

**Fig. 4.** Effect of energy resolution, spatial resolution, and timing resolution on TOF-PET resolution, sensitivity, contrast-to-noise ratio, and image quality. Gray highlights indicate baseline parameters: $k_E$ = 1 keV/switch, $\sigma_x$ = 1 mm, FWHM$_t$ = 500 ps. All Derenzo phantoms were run at 1/100 of a standard dose. a) FWHM (blue) and FWTM (orange) of PSF for varying $k_E$. b) Sensitivity for varying $k_E$, shown as total sensitivity (red / triangles), correct LORs (blue / squares), incorrect LORs (green / pentagons), and near-miss LORs (orange / circles). c) CNR for varying $k_E$ across different dot sizes for the Derenzo phantom. d) Highest-quality image for varying $k_E$ is obtained at $k_E$ = 1 keV/switch, at fixed $\sigma_x$ = 1 mm and FWHM$_t$ = 500 ps. e) FWHM (blue) and FWTM (orange) of PSF for varying $\sigma_x$. f) Sensitivity for varying $\sigma_x$, shown as total sensitivity (red / triangles), correct LORs (blue / squares), incorrect LORs (green / pentagons), and near-miss LORs (orange / circles). g) CNR for varying $\sigma_x$ across different dot sizes for the Derenzo phantom. h) Highest-quality image for varying $\sigma_x$ is obtained at $\sigma_x$ = 0.1 mm, at fixed $k_E$ = 1 keV/switch and FWHM$_t$ = 500 ps. i) FWHM (blue) and FWTM (orange) of PSF for varying FWHM$_t$. j) Sensitivity for varying FWHM$_t$, shown as total sensitivity (red / triangles), correct LORs (blue / squares), incorrect LORs (green / pentagons), and near-miss LORs

(orange / circles). k) CNR for varying FWHM$_t$ across different dot sizes for the Derenzo phantom. l) Highest-quality image for varying FWHM$_t$ is obtained at FWHM$_t$ = 50 ps, fixed $k_E$ = 1 keV/switch and σ$_x$ = 1 mm.

We tested scatter localization spatial resolutions, $\sigma_x$, of 0.1, 0.32, 1.0, and 3.2 mm (default: 1 mm). Fig. 4e-g show the effect of $\sigma_x$ on the image spatial resolution, the detector sensitivity, and the CNR for Derenzo rods. Smaller $\sigma_x$ directly improves image spatial resolution (Fig. 5e) but does not affect detector sensitivity (Fig. 5f) as the tested spatial resolutions are still small compared to typical distances between scatters. The measured CNR increases for smaller $\sigma_x$ as expected; the drop below 1 mm can be attributed to insufficient voxel sampling (Fig. 5g). The best image of the Derenzo phantom, shown in Fig. 4h, results from the best $\sigma_x$ (0.1 mm).

We tested timing resolutions, FWHM$_t$, of 50, 100, 500, 1000, and 5000 ps (default: 500 ps). Time resolution has little effect on the image spatial resolution of a point source because even the best timing resolutions translate to large spatial uncertainties compared to $\sigma_x$, as shown in Fig. 4i. Changing the time resolution has no effect on the detector sensitivity, as the Compton reconstruction algorithm does not currently use timing information, as shown in Fig. 4j. Timing does strongly affect CNR, as shown in Fig. 4k, with improved timing removing pileup of LORs outside of their origin location. This helps to remove a blurred background. The best image of the Derenzo phantom, shown in Fig. 4l, results from the best FWHM$_t$ (50 ps).

### E. Ultra-Low-Dose TOF-PET via Low-Z Detection

Potential clinical applications of the low-Z scanner were investigated by running simulations on a human phantom. Using the baseline detector parameters of $k_E$ = 1 keV/switch, $\sigma_x$ = 1 mm, FWHM$_t$ = 500 ps, we were able to generate clear images of the brain at ultra-low doses. The brain phantom is created using 4D Extended Cardiac-Torso (XCAT) phantom software[44], which produces voxelized patient geometries that can easily be simulated in TOPAS. The phantom had no anatomical abnormalities. Densities and atomic compositions of the tissues were modeled using data from the International Commission on Radiological Protection (ICRP)[58,59] and the International Commission on Radiation Units and Measurements (ICRU)[60]. Based on tissue-specific uptake ratios[61] and the standard dose as defined in *III. Methods*, we calculated the standard radiotracer dose in the brain to be 33 kBq/mL in the gray matter and 8.25 kBq/mL in the white matter. We imaged the brain phantom in the presence of the full XCAT body phantom. Simulations were run both without and with an added 20 mm diameter spherical lesion (99 kBq/mL) following Lee and co-workers[62].

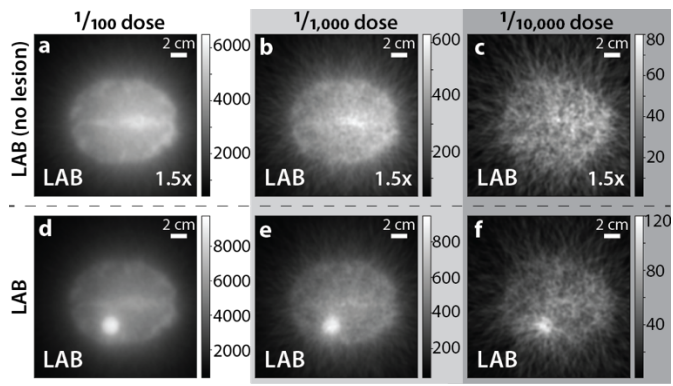

**Fig. 5.** Low-dose brain imaging, axial view. a)-c) Reconstructed images of the XCAT brain simulated using 1/100, 1/1,000, and 1/10,000 dose of 18F, respectively; each is shown at 1.5x contrast compared to the corresponding dose / image for d-f. d)-f) Reconstructed images of the XCAT brain + lesion, simulated using 1/100, 1/1,000, and 1/10,000 dose of 18F, respectively. Standard full-dose activity for white matter is taken to be 8.25 kBq/mL, gray matter is 33 kBq/mL, and the lesion is 99 kBq/mL; standard scan duration 10 min.

In Fig. 5, we present images of the brain at 1/100th, 1/1,000th, and 1/10,000th doses simulated in the low-Z detector. At 1/100th dose, the tumor, gray matter, and white matter can all be seen. At 1/1,000th dose, the tumor shape can still be seen. At 1/10,000th dose, the tumor is still apparent against the gray and white matter.

## V. Discussion

Here we have taken advantage of the TOPAS simulation framework for Geant4 to write a parametric simulation of a full-body TOF-PET detector based on an ionization-sensitive low-Z persistent recording medium and fast photodetectors to record the chain of Compton scatters from annihilation photons. The TOPAS framework facilitates high-fidelity physics simulations that separate the detector design from the underlying physics, so that detector variants and interchangeable phantoms can be easily tested. Using these simulations, we benchmarked the predicted performance of a low-Z detector for TOF-PET[27]. At reasonable detector specifications we also showed that such a detector could be used for ultra-low-dose brain imaging at 1% or even 0.1% of a standard radiotracer dose.

The projected capability to work with low doses stems from the high sensitivity of the low-Z detector. Given our detector geometry, the theoretical maximum sensitivity is 76.9% ($\frac{paired\ interactions}{photon\ pairs}$), so the 70% sensitivity predicted for the low-Z simulations in this work is nearly optimal. Since low-Z detection media are cheap relative to scintillation crystals, implementation of low-Z detectors with high stopping power is reasonably realistic. In addition to high stopping power, this low-Z design is optimized for measuring Compton scatters, which in low-Z materials dominates over photoelectric interactions. The high-Z crystals used in modern PET scanners generate a mix of Compton (~66%) and photoelectric interactions (~33%) while energy triggers are set to capture the photoelectric fraction.[31] There is no loss in sensitivity to the Compton fraction using methods optimized for low-Z.

The simulations presented in this work do not account for several factors that would influence design and efficacy of a real-world low-Z detector. For example, we assume that Compton scatters can be assigned to the correct chain, and that pairs of chains are correctly identified. Our simulations also do not account for any aspect of timing methodology; here shorter scan durations and lower radiotracer doses are treated as interchangeable, but these distinctions will become critical in the context of detection hardware and design choices. For example, pile-up cannot be fully explored without specific design of the detection electronics. We note, however, that pile-up scales with the square of the event rate, so that for this or any other low dose method, one would expect strongly reduced pile-up (for example, 1/100 dose translates to $10^{-4}$ pile-up). Moreover, the detector material proposed here is free of the radioactive background that contributes to pile-up in many scintillating crystal detectors. In addition, the combination of a large detection volume with precise timing provides a range of options for event separation as well as excitation and reset light patterning.

The approach described here could be further enhanced by using full timing and spatial information, including scintillation flashes, distances between scatters, and the correlation of the planes of scattering due to photon polarization effects; note that in this work only recoil electron energies $E$ and scattering angles were considered. Timing information was not included in the ordering algorithm as determining the time of scatters may require determining the ordering of the Compton chain. Additionally, if the directional trails of the recoil electrons can be visualized, the scattering locations could be determined much more precisely, and the recoil electron's initial trajectory could be included in the ordering algorithm. It might also be possible to infer the locations of in-patient scattering so that these detection events can also be used for image reconstruction. Without a sophisticated reconstruction algorithm, the predicted resolutions reported here are expected to underestimate the resolution performance of a fully developed system.[24] Direct reconstruction without filtered back-projection also leaves a $1/r^2$ falloff on all edges, limiting image sharpness. We expect that low-Z detector resolution could rapidly approach positron diffusion limits of tenths of a mm since scatter location should be readily measurable to a high degree of precision, particularly if any information about the recoil electron's trajectory is available.

Future prospects for this detection technique depend on development of an appropriate persistent low-Z medium to record Compton scatters. Among candidates, photo-switchable dyes in solution with a fast scintillator and energy-transfer-enhancing mediators, or another switchable two-state molecular system, may be desirable. A similar low-Z detection approach could be implemented for single photon emission computed tomography (SPECT) detection applications.

## VI. Conclusion

Ultra-low dose TOF-PET would profoundly impact both preclinical and clinical applications. For existing applications, higher sensitivity and resolution would allow smaller features such as early-stage tumor metastases to be identified, improving healthcare outcomes. Lowered radiation doses would be appropriate for a wider range of patients, including pediatric and pregnant populations. More PET scans could be performed for the same radioactive exposure, allowing more frequent and widespread use, for example for medical screening. Lower doses could also expand geographic access to TOF-PET. At 1/100 dose the radiotracer viability window could be extended by up to 6.5 half-lives; at 1/1000 dose it could be extended by as much as 10 half-lives, relaxing the required proximity to a cyclotron and other infrastructure. Our simulations show that ultra-low dose imaging could be done using a low-Z detection medium without stringent resolution requirements (1 keV / switch for energy resolution, 1 mm resolution of scatter location, and 500 ps FWHM timing resolution).

## Acknowledgements

For their exemplary software development and user support, we thank both Joseph Perl (TOPAS) and Paul Segars (XCAT). We thank Mary Heintz for essential computational system support. For financial support of undergraduate research from University of Chicago College, Physical Sciences Division, and Enrico Fermi Institute, we thank Steven Balla and Nichole Fazio, Michael Grosse, and Scott Wakely, respectively. We also thank the participants and organizers of the DOE / NIH workshop "Advancing Medical Care through Discovery in the Physical Sciences: Radiation Detection" (Jefferson Lab, April 2023) for their helpful discussions and feedback.

This work was supported in part by NIH R01EB026300 to P. J. La Riviere, a Neubauer Family Foundation award and a University of Chicago MRSEC seed grant (NSFDMR-2011854) to A. H. Squires, funding from the University of Chicago College, Physical Sciences Division and Enrico Fermi Institute to K. Domurat-Sousa, C. Poe, and J. F. Shida, and a University of Chicago Quad Undergraduate Research Scholarship to M. S. McDaniel.

# REFERENCES

1. Surti S, Karp JS. Update on latest advances in time-of-flight PET. *Physica Medica*. 2020;80:251-258. doi:10.1016/j.ejmp.2020.10.031

2. Cherry SR, Badawi RD, Karp JS, Moses WW, Price P, Jones T. Total-body imaging: Transforming the role of positron emission tomography. *Sci Transl Med*. 2017;9(381):eaaf6169. doi:10.1126/scitranslmed.aaf6169

3. Conti M, Bendriem B. The new opportunities for high time resolution clinical TOF PET. *Clin Transl Imaging*. 2019;7(2):139-147. doi:10.1007/s40336-019-00316-5

4. Moses WW. Fundamental limits of spatial resolution in PET. *Nuclear Instruments and Methods in Physics Research Section A: Accelerators, Spectrometers, Detectors and Associated Equipment*. 2011;648:S236-S240. doi:10.1016/j.nima.2010.11.092

5. Cherry SR. The 2006 Henry N. Wagner Lecture: Of mice and men (and positrons)--advances in PET imaging technology. *J Nucl Med*. 2006;47(11):1735-1745.

6. Moses WW. Recent advances and future advances in time-of-flight PET. *Nuclear Instruments and Methods in Physics Research Section A: Accelerators, Spectrometers, Detectors and Associated Equipment*. 2007;580(2):919-924. doi:10.1016/j.nima.2007.06.038

7. Surti S. Update on Time-of-Flight PET Imaging. *J Nucl Med*. 2015;56(1):98-105. doi:10.2967/jnumed.114.145029

8. Vaquero JJ, Kinahan P. Positron Emission Tomography: Current Challenges and Opportunities for Technological Advances in Clinical and Preclinical Imaging Systems. *Annu Rev Biomed Eng*. 2015;17(1):385-414. doi:10.1146/annurev-bioeng-071114-040723

9. Vandenberghe S, Moskal P, Karp JS. State of the art in total body PET. *EJNMMI Phys*. 2020;7(1):35. doi:10.1186/s40658-020-00290-2

10. Zhang X, Zhou J, Cherry SR, Badawi RD, Qi J. Quantitative image reconstruction for total-body PET imaging using the 2-meter long EXPLORER scanner. *Phys Med Biol*. 2017;62(6):2465-2485. doi:10.1088/1361-6560/aa5e46

11. Badawi RD, Shi H, Hu P, et al. First Human Imaging Studies with the EXPLORER Total-Body PET Scanner*. *J Nucl Med*. 2019;60(3):299-303. doi:10.2967/jnumed.119.226498

12. Conti M. Focus on time-of-flight PET: the benefits of improved time resolution. *Eur J Nucl Med Mol Imaging*. 2011;38(6):1147-1157. doi:10.1007/s00259-010-1711-y

13. Lecoq P, Morel C, Prior JO, et al. Roadmap toward the 10 ps time-of-flight PET challenge. *Phys Med Biol*. 2020;65(21):21RM01. doi:10.1088/1361-6560/ab9500

14. Van Sluis J, De Jong J, Schaar J, et al. Performance Characteristics of the Digital Biograph Vision PET/CT System. *J Nucl Med*. 2019;60(7):1031-1036. doi:10.2967/jnumed.118.215418

15. Lee MS, Cates JW, Gonzalez-Montoro A, Levin CS. High-resolution time-of-flight PET detector with 100 ps coincidence time resolution using a side-coupled phoswich configuration. *Phys Med Biol*. 2021;66(12):125007. doi:10.1088/1361-6560/ac01b5

16. Kwon SI, Ota R, Berg E, et al. Ultrafast timing enables reconstruction-free positron emission imaging. *Nat Photon*. 2021;15(12):914-918. doi:10.1038/s41566-021-00871-2

17. Ito M, Hong SJ, Lee JS. Positron emission tomography (PET) detectors with depth-of- interaction (DOI) capability. *Biomed Eng Lett*. 2011;1(2):70-81. doi:10.1007/s13534-011-0019-6

18. Pizzichemi M, Polesel A, Stringhini G, et al. On light sharing TOF-PET modules with depth of interaction and 157 ps FWHM coincidence time resolution. *Phys Med Biol*. 2019;64(15):155008. doi:10.1088/1361-6560/ab2cb0

19. Melcher CL. Scintillation crystals for PET. *J Nucl Med*. 2000;41(6):1051-1055.

20. Korzhik M, Fedorov A, Annenkov A, et al. Development of scintillation materials for PET scanners. *Nuclear Instruments and Methods in Physics Research Section A: Accelerators, Spectrometers, Detectors and Associated Equipment*. 2007;571(1-2):122-125. doi:10.1016/j.nima.2006.10.044

21. Moskal P, Niedźwiecki Sz, Bednarski T, et al. Test of a single module of the J-PET scanner based on plastic scintillators. *Nuclear Instruments and Methods in Physics Research Section A: Accelerators, Spectrometers, Detectors and Associated Equipment*. 2014;764:317-321. doi:10.1016/j.nima.2014.07.052

22. Reader AJ, Zaidi H. Advances in PET Image Reconstruction. *PET Clinics*. 2007;2(2):173-190. doi:10.1016/j.cpet.2007.08.001

23. Reader AJ, Corda G, Mehranian A, Costa-Luis CD, Ellis S, Schnabel JA. Deep Learning for PET Image Reconstruction. *IEEE Trans Radiat Plasma Med Sci*. 2021;5(1):1-25. doi:10.1109/TRPMS.2020.3014786

24. Spencer BA, Berg E, Schmall JP, et al. Performance Evaluation of the uEXPLORER Total-Body PET/CT Scanner Based on NEMA NU 2-2018 with Additional Tests to Characterize PET Scanners with a Long Axial Field of View. *J Nucl Med*. 2021;62(6):861-870. doi:10.2967/jnumed.120.250597

25. Oberla E, Genat JF, Grabas H, Frisch H, Nishimura K, Varner G. A 15GSa/s, 1.5GHz bandwidth waveform digitizing ASIC. *Nuclear Instruments and Methods in Physics Research Section A: Accelerators, Spectrometers, Detectors and Associated Equipment*. 2014;735:452-461. doi:10.1016/j.nima.2013.09.042

26. Frisch HJ, Oberla EJ, Kim HJ, Yeh M. Positron-emission tomography detector systems based on low-density liquid scintillators and precise time-resolving photodetectors. Published online November 20, 2018. https://www.osti.gov/biblio/1531371

27. Shida JF, Spieglan E, Adams BW, et al. Low-dose high-resolution TOF-PET using ionization-activated multi-state low-Z detector media. *Nuclear Instruments and Methods in Physics Research Section A: Accelerators, Spectrometers, Detectors and Associated Equipment*. 2021;1017:165801. doi:10.1016/j.nima.2021.165801

28. Aberle C, Elagin A, Frisch HJ, Wetstein M, Winslow L. Measuring directionality in double-beta decay and neutrino interactions with kiloton-scale scintillation detectors. *J Inst*. 2014;9(06):P06012-P06012. doi:10.1088/1748-0221/9/06/P06012

29. Adams BW, Elagin A, Frisch HJ, et al. Timing characteristics of Large Area Picosecond Photodetectors. *Nuclear Instruments and Methods in Physics Research Section A: Accelerators, Spectrometers, Detectors and Associated Equipment*. 2015;795:1-11. doi:10.1016/j.nima.2015.05.027

30. Adams BW, Attenkofer K, Bogdan M, et al. A Brief Technical History of the Large-Area Picosecond Photodetector (LAPPD) Collaboration. Published online March 6, 2016. Accessed May 5, 2023. http://arxiv.org/abs/1603.01843

31. Berger MJ, Hubbell JH, Seltzer SM, et al. XCOM: Photon Cross Section Database (version 1.5). Published online 2010. http://physics.nist.gov/xcom

32. Grignon C, Barbet J, Bardiès M, et al. Nuclear medical imaging using β+γ coincidences from 44Sc radionuclide with liquid xenon as detection medium. *Nuclear Instruments and Methods in Physics Research Section A: Accelerators, Spectrometers, Detectors and Associated Equipment*. 2007;571(1-2):142-145. doi:10.1016/j.nima.2006.10.048

33. Yoshida E, Tashima H, Nagatsu K, et al. Whole gamma imaging: a new concept of PET combined with Compton imaging. *Phys Med Biol*. 2020;65(12):125013. doi:10.1088/1361-6560/ab8e89

34. Perl J, Shin J, Schümann J, Faddegon B, Paganetti H. TOPAS: An innovative proton Monte Carlo platform for research and clinical applications: TOPAS: An innovative proton Monte Carlo platform. *Med Phys*. 2012;39(11):6818-6837. doi:10.1118/1.4758060

35. Faddegon B, Ramos-Méndez J, Schuemann J, et al. The TOPAS tool for particle simulation, a Monte Carlo simulation tool for physics, biology and clinical research. *Physica Medica*. 2020;72:114-121. doi:10.1016/j.ejmp.2020.03.019

36. Agostinelli S, Allison J, Amako K, et al. Geant4—a simulation toolkit. *Nuclear Instruments and Methods in Physics Research Section A: Accelerators, Spectrometers, Detectors and Associated Equipment*. 2003;506(3):250-303. doi:10.1016/S0168-9002(03)01368-8

37. Allison J, Amako K, Apostolakis J, et al. Geant4 developments and applications. *IEEE Trans Nucl Sci*. 2006;53(1):270-278. doi:10.1109/TNS.2006.869826

38. Allison J, Amako K, Apostolakis J, et al. Recent developments in Geant4. *Nuclear Instruments and Methods in Physics Research Section A: Accelerators, Spectrometers, Detectors and Associated Equipment*. 2016;835:186-225. doi:10.1016/j.nima.2016.06.125

39. Baró J, Sempau J, Fernández-Varea JM, Salvat F. PENELOPE: An algorithm for Monte Carlo simulation of the penetration and energy loss of electrons and positrons in matter. *Nuclear Instruments and Methods in Physics Research Section B: Beam Interactions with Materials and Atoms*. 1995;100(1):31-46. doi:10.1016/0168-583X(95)00349-5

40. Nuclear Energy Agency. *PENELOPE 2018: A Code System for Monte Carlo Simulation of Electron and Photon Transport: Workshop Proceedings, Barcelona, Spain, 28 January – 1 February 2019*. OECD; 2019. doi:10.1787/32da5043-en

41. PETRELAB 550-Q - C10-C13 Linear Alkylbenzene (LAB): Technical Data Sheet. Published online 2022. Accessed May 5, 2023. https://chemicals.cepsa.com/en/chemical-products/petrelab-550-q

42. Geant4 Collaboration. *User's Guide for Application Developers Using the Geant4 Toolkit v. 11.1 (Doc Rev7.0) - Appendix: Geant4 Material Database*. CERN; 2022. https://geant4-userdoc.web.cern.ch/UsersGuides/ForApplicationDeveloper/html/Appendix/materialNames.html

43. Derenzo SE. Monte Carlo simulations of time-of-flight PET with double-ended readout: calibration, coincidence resolving times and statistical lower bounds. *Phys Med Biol*. 2017;62(9):3828-3858. doi:10.1088/1361-6560/aa6862

44. Segars WP, Sturgeon G, Mendonca S, Grimes J, Tsui BMW. 4D XCAT phantom for multimodality imaging research: 4D XCAT phantom for multimodality imaging research. *Med Phys*. 2010;37(9):4902-4915. doi:10.1118/1.3480985

45. *NEMA Standards Publication NU 2-2018: Performance Measurements of Positron Emission Tomographs (PETS)*. National Electrical Manufacturers Association; 2018. http://wwww.nema.org

46. Boellaard R, O'Doherty MJ, Weber WA, et al. FDG PET and PET/CT: EANM procedure guidelines for tumour PET imaging: version 1.0. *Eur J Nucl Med Mol Imaging*. 2010;37(1):181-200. doi:10.1007/s00259-009-1297-4

47. Domurat-Sousa K, Poe CM, McDaniel MS, et al. TOPAS Simulation of a Low-Z TOF-PET Scanner (Version 1.0.0). Published online 2023. https://github.com/squireslab/low_z_pet_scanner

48. Matsuda K, Irie M. Diarylethene as a photoswitching unit. *Journal of Photochemistry and Photobiology C: Photochemistry Reviews*. 2004;5(2):169-182. doi:10.1016/S1389-5567(04)00023-1

49. Iwai R, Morimoto M, Irie M. Turn-on mode fluorescent diarylethenes: effect of electron-donating and electron-withdrawing substituents on photoswitching performance†. *Photochem Photobiol Sci*. 2020;19(6):783-789. doi:10.1039/d0pp00064g

50. Irie M. *Diarylethene Molecular Photoswitches: Concepts and Functionalities*. Wiley-VCH; 2021.

51. DeBenedetti S, Cowan CE, Konneker WR, Primakoff H. On the Angular Distribution of Two-Photon Annihilation Radiation. *Phys Rev*. 1950;77(2):205-212. doi:10.1103/PhysRev.77.205

52. Levin CS, Hoffman EJ. Calculation of positron range and its effect on the fundamental limit of positron emission tomography system spatial resolution. *Phys Med Biol*. 1999;44(3):781-799. doi:10.1088/0031-9155/44/3/019

53. Blanco A. Positron Range Effects on the Spatial Resolution of RPC-PET. In: *2006 IEEE Nuclear Science Symposium Conference Record*. IEEE; 2006:2570-2573. doi:10.1109/NSSMIC.2006.354433

54. Geant4 Collaboration. *User's Guide: Physics Reference Manual for Geant4 Toolkit v. 11.1 (Doc Rev7.0)*. CERN; 2022. https://geant4-userdoc.web.cern.ch/UsersGuides/PhysicsReferenceManual/html/index.html

55. Haynes WM, ed. *CRC Handbook of Chemistry and Physics*. 0 ed. CRC Press; 2014. doi:10.1201/b17118

56. Klein O, Nishina Y. The Scattering of Light by Free Electrons according to Dirac's New Relativistic Dynamics. *Nature*. 1928;122(3072):398-399. doi:10.1038/122398b0

57. Derenzo SE, Budinger TF, Cahoon JL, Huesman RH, Jackson HG. High Resolution Computed Tomography of Positron Emitters. *IEEE Trans Nucl Sci*. 1977;24(1):544-558. doi:10.1109/TNS.1977.4328738

58. Clement CH, International Commission on Radiological Protection, eds. *Adult Reference Computational Phantoms: Joint ICRP/ICRU Report*. Elsevier; 2009.

59. Kim CH, Yeom YS, Petoussi-Henss N, et al. *Adult Mesh-Type Reference Computational Phantoms*. SAGE; 2020.

60. International Commission on Radiation Units and Measurements. *Photon, Electron, Proton and Neutron Interaction Data for Body Tissues*. International Commission on Radiation Units and Measurements; 1992. https://www.osti.gov/biblio/7114520

61. Verwer EE, Golla SSV, Kaalep A, et al. Harmonisation of PET/CT contrast recovery performance for brain studies. *Eur J Nucl Med Mol Imaging*. 2021;48(9):2856-2870. doi:10.1007/s00259-021-05201-w

62. Lee J, Oya S, Sade B. Benefits and limitations of diameter measurement in the conservative management of meningiomas. *Surg Neurol Int*. 2011;2(1):158. doi:10.4103/2152-7806.89857

# Simulation of a Compton-based detector
# for low-dose high-resolution time-of-flight PET
# Supporting Information

Kepler Domurat-Sousa[1,2], Cameron M. Poe[1], Maya S. McDaniel[2], Eric Spieglan[1], Joao F. Shida[1], Evan Angelico[1], Bernhard W. Adams[3], Patrick J. La Riviere[4], Henry J. Frisch[1], and Allison H. Squires[2,5]

6.  Enrico Fermi Institute, University of Chicago, Chicago, IL USA
7.  Pritzker School of Molecular Engineering, University of Chicago, Chicago, IL, USA
8.  Quantum Optics Applied Research, Naperville, IL, USA
9.  Department of Radiology at the Pritzker School of Medicine, University of Chicago, Chicago, IL, USA
10. Institute for Biophysical Dynamics at the University of Chicago, IL, USA

## VII. CONTRAST-TO-NOISE RATIO (CNR)

This section provides additional details of the CNR calculations performed for the Derenzo phantom. Fig. 1a shows the rod numbering scheme referred to throughout this section. Rods are numbered from largest to smallest, inside to outside. Fig. 1b and 1c show the signal and background masks, respectively, applied here to a reconstructed image from the Compton based simulation.

Based on the data from these masked regions and from Fig. 4 of the main paper, there are apparent spatial biases in the resulting CNR calculations, as illustrated by Fig. 1 d-f. The outer rods have significantly lower CNR values because the outer edges of their background regions are generally darker than rod backgrounds closer to the middle of the image, which makes for both higher contrast (Fig. 1d) and higher noise (Fig. 1e). However, the difference in the noise between these outer rods and the others is greater than differences in contrast, which creates a lower overall CNR value. This is because low spatial frequency brightness changes dominate the noise, with the background gradually lightening as it moves from the outer edge of the phantom toward the center, which occurs due to the direct image reconstruction approach which creates $1/r^2$ falloff.

Systematic spatial variations are also responsible for the drop in average CNR values for the largest rod size as shown in Fig. 1f. Fig. 1c reveals that the background around the largest rods also exhibit the $1/r^2$ effect near the rod edges, appearing dimmer in between. This effect is most pronounced for the largest rods, where the spacing is also largest, inflating the standard deviation and lowering the CNR.

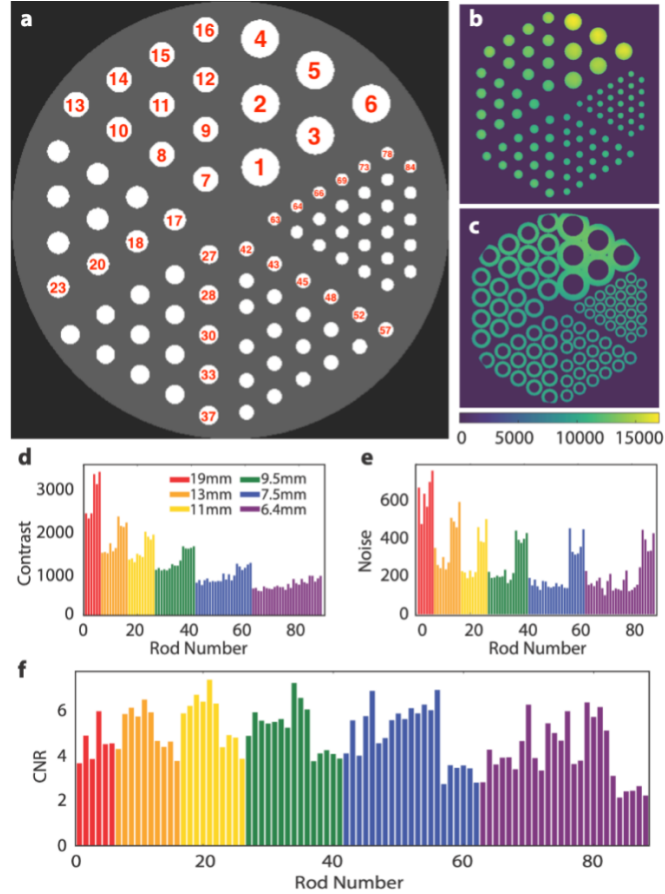

**Fig. 1.** CNR masks and individual rod CNR values for the Derenzo phantom. All panels refer to the simulation for LAB with baseline parameters. a) The numbering scheme for the rods as later represented in d-f. b) The mask used to determine signal of each rod, covering the entire diameter of the rod as given by the truth information. c) A mask depicting the areas used to calculate noise of each rod, covering a ring from 1.2 to 1.9 times the radius of the rod, but excluding anything within ten pixels of the phantom's true edge. d) Contrast values for each rod, calculated by subtracting the average brightness of the background ring from the average brightness of the rod. e) Noise values for each rod, calculated with the standard deviation of the brightness of each background ring. f) CNR for each rod, calculated by dividing the contrast by the noise per Eqn. 6 in the main text.

## VIII.    NEMA NU2-2018 Sensitivity

The NEMA NU2-2018 sensitivity source was modeled as a line source 70 cm long with a hollow cylinder of aluminum around it of varying thickness. In each run 1,000,000 positrons were created in the line source, and the number of LORs created was recorded. The counts were then fit to an exponential to determine the sensitivity in the no-attenuator limit.

The NEMA NU2-2018 sensitivity number is reported to provide a comparable value between this simulation and other PET scanner concepts. For comparisons within the paper of varying detector parameters a point source with zero attenuation was used. This allowed for calculation of the sensitivity from a single simulation run, and allowed for direct measurement of the near-miss and miss LORs.

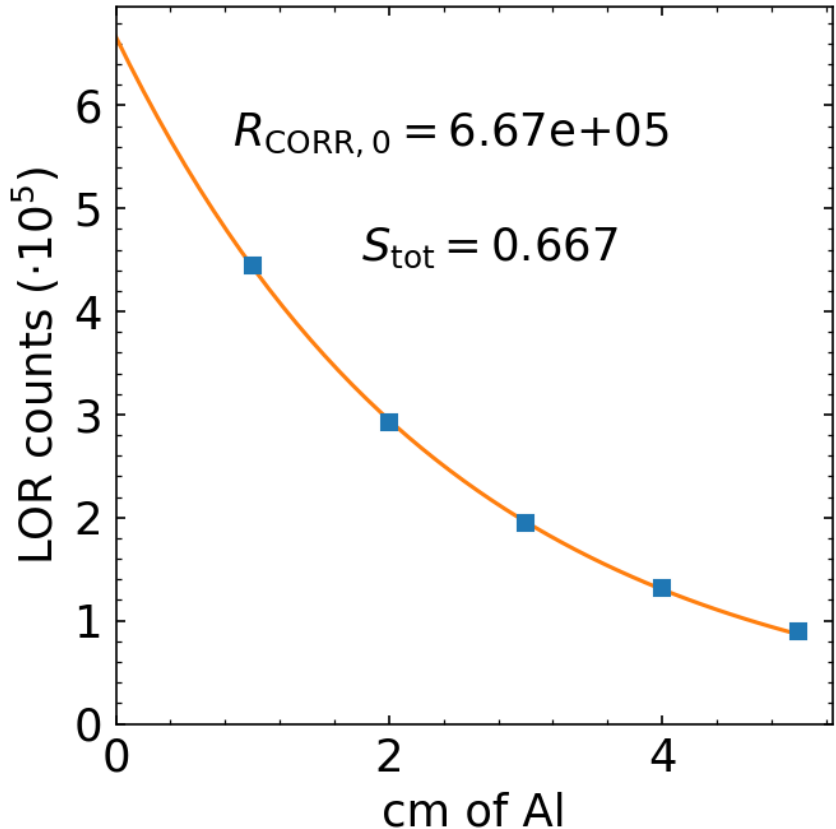

**Fig. 2.** The counted LORs for various attenuating thicknesses of Aluminum around a 70 cm line source. Fitting to an exponential gives a no-attenuation sensitivity of 66.7%

## IX.   REJECTION OF IN-PATIENT SCATTERING

In scintillation-based TOF-PET detectors, rejection of in-patient scattering (IPS) depends on identifying events with one or more photons undergoing IPS by the energy of the scattered photon being less than the initial 511 keV. The rejection consequently crucially depends on the energy resolution of the scintillation and photodetector systems.

In a Compton-based detector the rejection of IPS is an outcome of the ordering of the Compton chain of scatters. In a switchillator-based Compton detector, the energy is measured by the number of switched molecules at each scattering site. Energy resolution then enters through the kinematic constraints of the Compton 2-body process during the process of ordering the scatters by energy and angle, and is not used directly in IPS rejection.

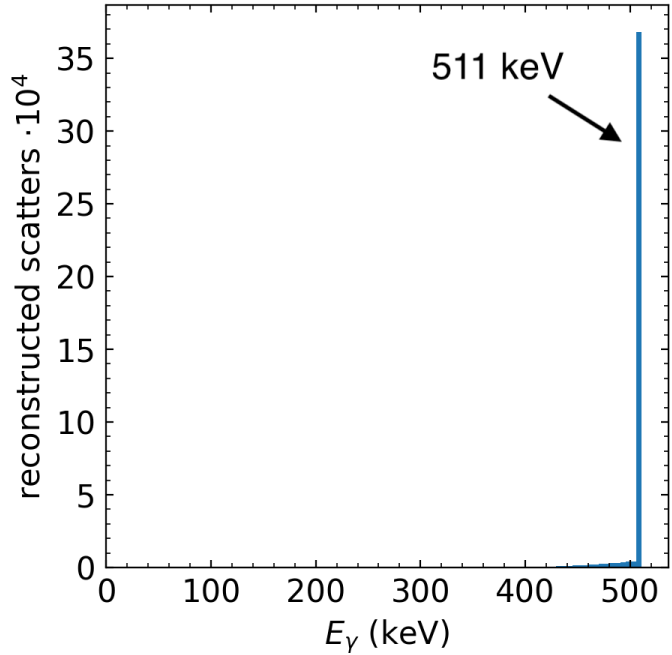

**Fig. 3.** The energy of photons in pair events that pass the selection criteria for making an image of a point source of 1,000,000 positrons in a cylinder of water the size of the Derenzo phantom (linear scale). Note that 86.9% are at the unscattered energy of emission of 511 keV.

However, in a simulation study one can ask *post facto* what are the ground truth energies of the photons comprising pair events that pass the criteria for inclusion in an image (the "effective energy resolution" of the IPS rejection). This represents the energy resolution of a scintillator-based detector that would be required to provide the same rejection, which in this study is a rejection of 90% of pair events with one or more scattered photons.

Figures 3 and 4 show the energies of both photons in events that pass the image criteria for a point source of 1,000,000 positrons in a cylinder of water the size of the Derenzo phantom, on a linear scale and a logarithmic scale, respectively. The selection of the first scatter in a reconstructed Compton chain thus rejects in-patient scattering more cleanly than with an energy cut; the equivalent energy resolution, shown by the distributions in the plots, is far better than the stochastic photometric processes in calorimetry can provide.

The Compton chain algorithm calculates the predicted energy of the incident photon that undergoes Compton scattering in the detector in two different ways. The first is a calculation using the geometry and deposited energies of the scattered photon and electrons. This gives a value $E^{Comp} \pm \delta E^{Comp}$ based only on the Compton scattering constraints. A second value $E^{chain} \pm \delta E^{chain}$ is calculated using

$$E_i^{chain} = 511 \text{ keV} - \sum_{j=1}^{i} E_j^{dep}$$

where $E_j^{dep}$ is the energy deposited at scatter $j$.

In instances where the geometry is incorrect (typically that the wrong scatter ordering has been selected) the two values $E^{Comp}$ and $E^{chain}$ will differ, with their difference increasing farther down the chain. This leads to the ordering being rejected.

In instances where the photon entering the detector in-patient scattered (IPS), the assumption that the incoming photon was at 511 keV will be incorrect. This causes deviations between $E^{Comp}$ and $E^{chain}$. This deviation leads to the Compton chains being rejected. In IPS cases there is usually no possible chain ordering that satisfies the constraints and so all possible line-of-response get cut.

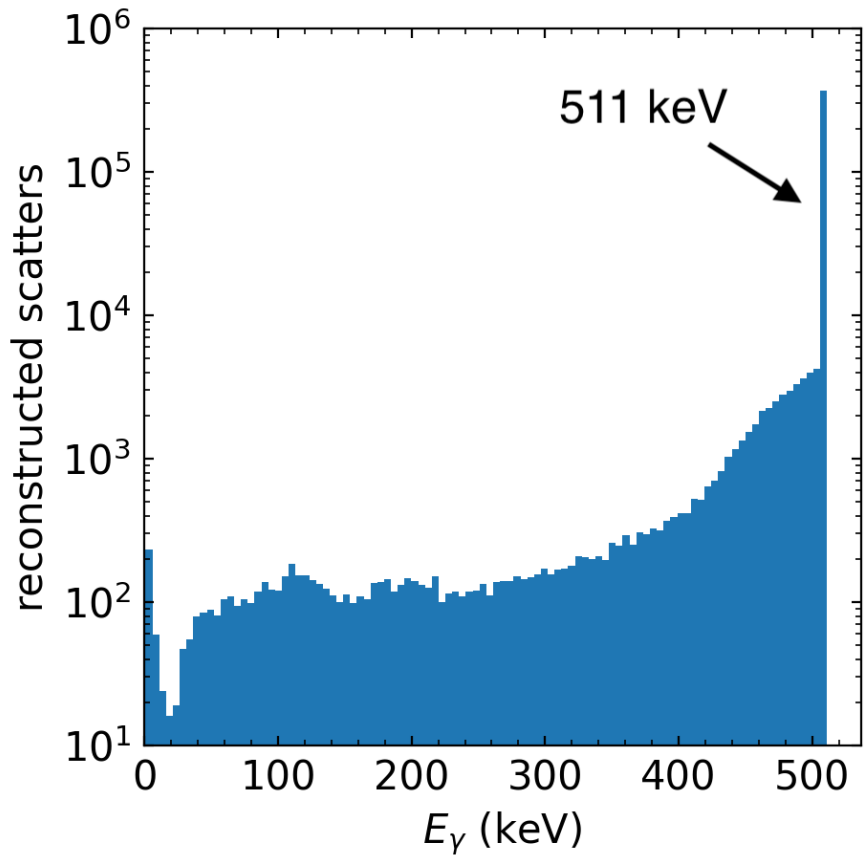

**Fig. 4.** The energy of photons in pair events that pass the selection criteria for making an image of a point source of 1,000,000 positrons in a cylinder of water the size of the Derenzo phantom (logarithmic scale). Note that 86.9% are at the unscattered energy of emission of 511 keV.